\documentclass[aps,pra,twocolumn,showpacs,superscriptaddress]{revtex4-1}

\usepackage{amsfonts,mathrsfs,amsmath,amsthm,amssymb}
\usepackage{bm,times}
\usepackage{graphicx,epsfig}
\usepackage[colorlinks=true,breaklinks=true,linkcolor=blue,citecolor=blue,urlcolor=blue]{hyperref}

\def \non {\nonumber}

\newcommand{\be}{\begin{eqnarray}}
\newcommand{\ee}{\end{eqnarray}}

\makeatletter
    
    \newcommand{\Rmnum}[1]{\expandafter\@slowromancap\romannumeral #1@}
\makeatother

\begin{document}
\title{Classification and purification for the independent quantum channel through quantum error-correction}
\author{Long Huang}
\email{huangl@sicnu.edu.cn}
\affiliation{Coll Phys \& Elect Engn, Sichuan Normal University, Chengdu 610101, China}
\author{Xiaohua Wu}
\email{wxhscu@scu.edu.cn}
\affiliation{College of Physical Science and Technology, Sichuan University, Chengdu 610064, China}
\author{Tao Zhou}
\email{taozhou@swjtu.edu.cn}
\affiliation{Department of Applied Physics, School of Physical Science and Technology, Southwest Jiaotong University, Chengdu 611756, China}

\begin{abstract}
The essence of quantum error-correction is to use redundant Hilbert space to identify and correct errors, and the channel fidelity of the quantum channel does not affect which errors can be identified and corrected. Based on this, it is found that quantum error-correction can be used to classify the independent quantum channel into 5 types, and 4 of the 5 types can be purified. It is found in quantum error-correction, the decoherence of quantum state may be related to the degree of identification for the state under quantum noise, and the results of this work confirmed that the degree of purity of quantum channel determines its ability to retain the quantum property of the quantum state, not the fidelity. In this work, the identification of the independent Pauli channels by quantum error-correction is demonstrated.
\end{abstract}

\date{\today}

\maketitle

\section{Introduction}

In quantum computation and communication, quantum error-correction (QEC) developed from classic schemes to preserving coherent states from noise and other unexpected interactions. QEC codes are introduced as active error-correction. Another way, passive error-avoiding techniques contain decoherence-free subspaces~\cite{Duan,Lidar,Zanardi} and noiseless subsystem~\cite{KandLV,Zanardi2,Kempe}. Recently, it has been proven that both the active and passive QEC methods can be unified~\cite{Kribs,Poulin 05,Kribs2}. The main research objects in QEC include quantum noise channel and error-correction protocol~\cite{Nielsen}. In order to applying QEC in practical situations without measuring the quantum channels, the error-correction protocol we adopted had better be robust to most noise channels.

In this work, we consider using the Pauli channel to represent the independent noise channel. Because in recent works, it is found the Pauli channel is the approaching target of the effective channel after QEC~\cite{Huang,J. J. Wallman,S. J. Beale,C.C2017,E. Huang,Huang2}, which makes it relatively easier to analyze the evolution of the noise channel. On the other hand, we expand the range of channel fidelity for the Pauli channel as $\{\lambda_{II}=p_1,\lambda_{XX}=p_2,\lambda_{YY}=p_3,\lambda_{ZZ}=p_4,p_1+p_2+p_3+p_4=1\}$, which is reasonable not only because the real situation exists but also the logical $X,Y,Z$ operators on the logical state will introduce the situation in main system qubit.

Meanwhile, we fix the measurement-free QEC scheme~\cite{Huang,Huang2,Bennett}, which consists of encoding operation $\mathcal{U}$, the noise evolution $\Lambda$, and decoding operation $\mathcal{U}^{-1}$. The unitary process $\mathcal{U}$ represents encoding and its associated process $\mathcal{U}^{-1}$ represents decoding with recovery. The number of logical states that can be identified is $2^{n-1}$, and the number of errors that can be recovered is at least $2^{n-1}$ (because part of errors result in the same logical state), where $n$ is the number of qubits in quantum code. The value of channel fidelity does not affect which errors can be identified and corrected, and it just affects the channel fidelity of the effective channel.

Through analog calculation, it is found that QEC not only can be used for noise suppression, but also for classification and purification of the independent quantum channel. The independent quantum channel can be classified into 5 types, and 4 of the 5 types can be purified with QEC. If the dominant role of one channel is similar to the $I$ operator, and this channel can be purified with QEC, it is called an $I$-type channel. Meanwhile, the definitions of $X$-type, $Y$-type, and $Z$-type are similar to the $I$-type channel. On the other hand, the channel that can not be purified is called the $T$-type (Trivial-type) channel. After concatenated QEC, the effective channels will converge to one of the five types channel, which is easy to distinguish through quantum process tomography~\cite{Gilchrist,Emerson,Knill 08,Bendersky,Magesan 11,Magesan 12,X.-H. Wu}, this provides a method to trace to the source of the quantum channel.

Based on the analysis of results, there are two methods to realize the logical Pauli operators in the QEC scheme. The one is applying Pauli operators on the main system qubit before encoding, which will introduce correlated noise. The another one is applying logical Pauli operators after encoding or before decoding, which will introduce independent noise. Because of the processes are not ideal, the effects introduced by the two methods are different. It is found that the difference of the two methods can be related to the difference between correlated noise and independent noise in the QEC scheme, which corresponding to the degree of identification of the correlated noise and the independent noise. Furthermore, we consider that the decoherence of quantum state may be related with the degree of identification for the state under quantum noise, and the results of this work confirmed that the degree of purity of quantum channel determines its ability to retain the quantum property of the quantum state, not the fidelity.

Although using the purity of the state as a performance metric rather than the fidelity sounds very similar to the Pauli-tracking approach, we would like to point out that the QEC scheme has actually been followed the rule without additional operations.

The content of the present work is organized as follows. In Sec.~\ref{ii}, we process the tomography of the Pauli channel with seven error-correction codes: $3$-qubit code, $7$-qubit Steane code~\cite{Steane}, $9$-qubit Shor code~\cite{Shor}, two new $9$-qubit codes~\cite{Huang3}, and the two $5$-qubit codes~\cite{Bennett,Laflamme}. In Sec.~\ref{iii}, we analyze the difference between correlated noise and independent noise in the QEC scheme by discussing the realization of logical Pauli operators. In Sec.~\ref{iv}, we end this work with some remarks and discussion.

\section{Computations}
\label{ii}

In this section, we study the classification and purification of the independent quantum channel with QEC. Based on the specific performances, the error-correction codes can be divided into two categories. One of them can normally purify the independent quantum channel, which contains $3$-qubit code, $7$-qubit Steane code, and the $5$-qubit code in Ref.~\cite{Bennett}. The other category will introduce some fixed logical operators during the purification of the independent quantum channel, which contains $9$-qubit Shor code and the $5$-qubit code in Ref.~\cite{Laflamme}. It is found that the eigenvalues of $I,X,Y,Z$ in effective channel are affected by the structure of the code and the choice of correctable errors.

The logical states of $3$-qubit code are $|0_\mathcal{L}\rangle=|000\rangle$ and $|1_\mathcal{L}\rangle=|111\rangle$, and the stabilizer generators are $g_1=Z_1Z_2$ and $g_2=Z_2Z_3$. The process matrix $\lambda$ of bit-flip noise channel is $\{\lambda_{II}=p_1,\lambda_{XX}=p_2,p_1+p_2=1\}$, the correctable errors usually are set as $I$, $X_1$, $X_2$, and $X_3$. After one-level QEC, the effective channel is obtained
\be
\label{e1}
\tilde{\lambda}=\{\tilde{\lambda}_{II}=p^3_1+3p^2_1p_2,\tilde{\lambda}_{XX}=p^3_2+3p_1p^2_2\},
\ee
the symmetry of this formula indicates that the effective channel is approaching identity operation channel $I$ ($p_1>\frac{1}{2}$) or bit-flip operation channel $X$ ($p_1<\frac{1}{2}$) through concatenated QEC.

The $7$-qubit Steane code has six stabilizer generators,
\be
g_1&=&Z_4Z_5Z_6Z_7,g_2=X_4X_5X_6X_7,\non\\
g_3&=&Z_2Z_3Z_6Z_7,g_4=X_2X_3X_6X_7,\non\\
g_5&=&Z_1Z_3Z_5Z_7,g_6=X_1X_3X_5X_7.\non
\ee
The logical state $|0_\mathcal{L}\rangle$ is generated by applying all combinations from $\{I,g_2,g_4,g_6\}$ on $|0^{\otimes7}\rangle$, and the logical state $|1_\mathcal{L}\rangle$ is obtained by applying $X^{\otimes7}$ on $|0_\mathcal{L}\rangle$~\cite{Huang3}. For general Pauli channel $\{\lambda_{II}=p_1,\lambda_{XX}=p_2,\lambda_{YY}=p_3,\lambda_{ZZ}=p_4,p_1+p_2+p_3+p_4=1\}$, the set of correctable errors consists of the identity operator $I$, all the weight-one Pauli operators and $42$ weight-two Pauli operators. After one-level QEC, the effective channel is obtained
\be
\label{e2}
\tilde{\lambda}=\{\tilde{\lambda}_{II}=p^7_1+7p^6_1(p_2+p_3+p_4)+p^5_1(3p_2p_3\non\\
+19p_3p_4+20p_2p_4)+O(p^{i,i=0,1,..5}_1),\non\\
\tilde{\lambda}_{XX}=p^7_2+7p^6_2(p_1+p_3+p_4)+p^5_2(3p_1p_4\non\\
+19p_3p_4+20p_1p_3)+O(p^{i,i=0,1,..5}_2),\non\\
\tilde{\lambda}_{YY}=p^7_3+7p^6_3(p_1+p_2+p_4)+p^5_3(3p_1p_4\non\\
+19p_1p_2+20p_2p_4)+O(p^{i,i=0,1,..5}_3),\non\\
\tilde{\lambda}_{ZZ}=p^7_4+7p^6_4(p_1+p_2+p_3)+p^5_4(3p_2p_3\non\\
+19p_1p_2+20p_1p_3)+O(p^{i,i=0,1,..5}_4)\}.
\ee
Here, $O(p^{i}_j)$ is the low-order term of $p^{i}_j$ ($j=1,2,3,4,i=0,1,2,3,4,5$). The identity operator $I$ and all the weight-one Pauli operators of correctable errors result in the same coefficient items for $\tilde{\lambda}_{II}, \tilde{\lambda}_{XX}, \tilde{\lambda}_{YY}, \tilde{\lambda}_{ZZ}$, which is $p^7+7p^6(1-p)$. Taking only the dominant terms when $p_j$ is close to $1$, the eigenvalues of $I,X,Y,Z$ after QEC have the same form at the minimum, written as $21(1-p)^5p^2+7(1 - p)^4p^3+28(1-p)^3p^4+7(1-p)p^6+p^7$. It indicates that when p is above $0.935404$, the quantum channel can be purified.

The $5$-qubit code (I) from Ref.~\cite{Bennett} has four stabilizer generators, which are
\be
g_1&=&X_1Z_2Z_3X_4,g_2=X_2Z_3Z_4X_5,\non\\
g_3&=&X_1X_3Z_4Z_5,g_4=Z_1X_2X_4Z_5.\non
\ee
The logical state $|0_\mathcal{L}\rangle$ is generated by applying all combinations from $\{I,g_1,g_2,g_3,g_4\}$ on $|0^{\otimes5}\rangle$, and the logical state $|1_\mathcal{L}\rangle$ is obtained by applying $X^{\otimes5}$ on $|0_\mathcal{L}\rangle$~\cite{Huang3}. For general Pauli channel $\{\lambda_{II}=p_1,\lambda_{XX}=p_2,\lambda_{YY}=p_3,\lambda_{ZZ}=p_4,p_1+p_2+p_3+p_4=1\}$, the set of correctable errors consists of the identity operator $I$, and all the weight-one Pauli operators. After one-level QEC, the effective channel is obtained
\be
\label{e3}
\tilde{\lambda}=\{
\tilde{\lambda}_{II}=p^5_1+5p^4_1(p_2+p_3+p_4)+10p^2_1(p^2_2p_3+p_2p^2_3\non\\
+p^2_2p_4+p_2p^2_4+p^2_3p_4+p_3p^2_4)+O(p^{i,i=0,1}_1),\non\\
\tilde{\lambda}_{XX}=p^5_2+5p^4_2(p_1+p_3+p_4)+10p^2_2(p^2_1p_3+p_1p^2_3\non\\
+p^2_1p_4+p_1p^2_4+p^2_3p_4+p_3p^2_4)+O(p^{i,i=0,1}_2),\non\\
\tilde{\lambda}_{YY}=p^5_3+5p^4_3(p_1+p_2+p_4)+10p^2_3(p^2_1p_2+p_1p^2_2\non\\
+p^2_1p_4+p_1p^2_4+p^2_2p_4+p_2p^2_4)+O(p^{i,i=0,1}_3),\non\\
\tilde{\lambda}_{ZZ}=p^5_4+5p^4_4(p_1+p_2+p_3)+10p^2_4(p^2_1p_2+p_1p^2_2\non\\
+p^2_1p_3+p_1p^2_3+p^2_2p_3+p_2p^2_3)+O(p^{i,i=0,1}_4)\}.
\ee
Here, $O(p^{i}_j)$ is the low-order term of $p^{i}_j$ ($j=1,2,3,4,i=0,1$). For the 5-qubit code, there is only one choice of correctable errors, which result in similar coefficient items for $\tilde{\lambda}_{II}, \tilde{\lambda}_{XX}, \tilde{\lambda}_{YY}, \tilde{\lambda}_{ZZ}$. Taking only the dominant terms when $p_j$ is close to $1$, the eigenvalues of $I,X,Y,Z$ after QEC have the same form at the minimum, written as $5 p^4(1 - p)  + p^5$. It indicates that when p is above $0.868877$, the quantum channel can be purified.

The logical states of the two $9$-qubit codes in Ref.~\cite{Huang3} are,
\be
|0_\mathcal{L}\rangle
&=&\frac{1}{2}[|000000000\rangle+|111111000\rangle\non\\
&&+|000111111\rangle+|111000111\rangle],\non\\
|1_\mathcal{L}\rangle
&=&\frac{1}{2}[|111111111\rangle+|000000111\rangle\non\\
&&+|111000000\rangle+|000111000\rangle],\non
\ee
and
\be
|0_\mathcal{L}\rangle=\frac{1}{8}[|000\rangle+|011\rangle+|101\rangle+|110\rangle]^{\bigotimes3},\non\\
|1_\mathcal{L}\rangle=\frac{1}{8}[|111\rangle+|100\rangle+|010\rangle+|001\rangle]^{\bigotimes3}.\non
\ee
One set generators are $Z_1Z_2,Z_2Z_3$, $Z_4Z_5,Z_5Z_6$, $Z_7Z_8,Z_8Z_9$, $X_1X_2X_3X_4X_5X_6$, and $X_4X_5X_6X_7X_8X_9$, and the other are $X_1X_2,X_2X_3$, $X_4X_5,X_5X_6$, $X_7X_8,X_8X_9$, $Z_1Z_2Z_3Z_4Z_5Z_6$, and $Z_4Z_5Z_6Z_7Z_8Z_9$. For general Pauli channel $\{\lambda_{II}=p_1,\lambda_{XX}=p_2,\lambda_{YY}=p_3,\lambda_{ZZ}=p_4,p_1+p_2+p_3+p_4=1\}$, the set of correctable errors consists of the identity operator $I$, all the weight-one Pauli operators, and other $228$ weight-two to weight-three Pauli operators. After one-level QEC, the effective channel has the same dominant terms,
\be
\label{e10}
\tilde{\lambda}=\{
\tilde{\lambda}_{II}=p^9_1+9p^8_1(p_2+p_3+p_4)+9p^7_1(p^2_4+2p_3p_4\non\\
+3p^2_2+3p^2_3+6p_2p_3+8p_2p_4)+O(p^{i,i=0,1,..6}_1),\non\\
\tilde{\lambda}_{XX}=p^9_2+9p^8_2(p_1+p_3+p_4)+9p^7_2(p^2_3+2p_3p_4\non\\
+3p^2_1+3p^2_4+6p_1p_4+8p_1p_3)+O(p^{i,i=0,1,..6}_2),\non\\
\tilde{\lambda}_{YY}=p^9_3+9p^8_3(p_1+p_2+p_4)+9p^7_3(p^2_2+2p_1p_2\non\\
+3p^2_1+3p^2_4+6p_1p_4+8p_2p_4)+O(p^{i,i=0,1,..6}_3),\non\\
\tilde{\lambda}_{ZZ}=p^9_4+9p^8_4(p_1+p_2+p_3)+9p^7_4(p^2_1+2p_1p_2\non\\
+3p^2_2+3p^2_3+6p_2p_3+8p_1p_3)+O(p^{i,i=0,1,..6}_4)\}.
\ee
Here, $O(p^{i}_j)$ is the low-order term of $p^{i}_j$ ($j=1,2,3,4,i=0,1,2,3,4,5,6$). The identity operator $I$ and all the weight-one Pauli operators of correctable errors result in the same coefficient items for $\tilde{\lambda}_{II}, \tilde{\lambda}_{XX}, \tilde{\lambda}_{YY}, \tilde{\lambda}_{ZZ}$, which is $p^9+9p^8(1-p)$. Taking only the dominant terms when $p_j$ is close to $1$, the eigenvalues of $I,X,Y,Z$ after QEC have the same form at the minimum, written as $27 (1 - p)^7 p^2 + 27 (1 - p)^6 p^3 + 99 (1 - p)^5 p^4 + 27 (1 - p)^4 p^5 + 57 (1 - p)^3 p^6 + 9 (1 - p)^2 p^7 + 9 (1 - p) p^8 + p^9$. It indicates that when p is above $0.950149$, the quantum channel can be purified.

The logical states of $9$-qubit Shor code $|0_\mathcal{L}\rangle$ is generated by all combinations from $\{|000\rangle, |111\rangle\}$, and the logical state $|1_\mathcal{L}\rangle$ is obtained by applying $Z^{\otimes9}$ on $|0_\mathcal{L}\rangle$. The eight generators of $9$-qubit Shor code are $g_1=Z_1Z_2$, $g_2=Z_2Z_3$, $g_3=Z_4Z_5$, $g_4=Z_5Z_6$, $g_5=Z_7Z_8$, $g_6=Z_8Z_9$, $g_7=X_1X_2X_3X_4X_5X_6$ and $g_8=X_4X_5X_6X_7X_8X_9$. For general Pauli channel $\{\lambda_{II}=p_1,\lambda_{XX}=p_2,\lambda_{YY}=p_3,\lambda_{ZZ}=p_4,p_1+p_2+p_3+p_4=1\}$, the set of correctable errors consists of the identity operator $I$, all the weight-one Pauli operators, and other $228$ weight-two to weight-three Pauli operators. After one-level QEC, the effective channel is obtained
\be
\label{e4}
\tilde{\lambda}=\{
\tilde{\lambda}_{II}=p^9_1+9p^8_1(p_2+p_3+p_4)+9p^7_1(p^2_4+2p_3p_4\non\\
+3p^2_2+3p^2_3+6p_2p_3+8p_2p_4)+O(p^{i,i=0,1,..6}_1),\non\\
\tilde{\lambda}_{XX}=p^9_4+9p^8_4(p_1+p_2+p_3)+9p^7_4(p^2_1+2p_1p_2\non\\
+3p^2_2+3p^2_3+6p_2p_3+8p_1p_3)+O(p^{i,i=0,1,..6}_4),\non\\
\tilde{\lambda}_{YY}=p^9_3+9p^8_3(p_1+p_2+p_4)+9p^7_3(p^2_2+2p_1p_2\non\\
+3p^2_1+3p^2_4+6p_1p_4+8p_2p_4)+O(p^{i,i=0,1,..6}_3),\non\\
\tilde{\lambda}_{ZZ}=p^9_2+9p^8_2(p_1+p_3+p_4)+9p^7_2(p^2_3+2p_3p_4\non\\
+3p^2_1+3p^2_4+6p_1p_4+8p_1p_3)+O(p^{i,i=0,1,..6}_2)\}.
\ee
Here, $O(p^{i}_j)$ is the low-order term of $p^{i}_j$ ($j=1,2,3,4,i=0,1,2,3,4,5,6$). The identity operator $I$ and all the weight-one Pauli operators of correctable errors result in the same coefficient items for $\tilde{\lambda}_{II}, \tilde{\lambda}_{XX}, \tilde{\lambda}_{YY}, \tilde{\lambda}_{ZZ}$, which is $p^9+9p^8(1-p)$. Meanwhile, we notice that the eigenvalue of $X$ is the main determined by the initial eigenvalue of $Z$, and the eigenvalue of $Z$ is main determined by the initial eigenvalue of $X$, which may result from the structure of the $9$-qubit Shor code. Taking only the dominant terms when $p_j$ is close to $1$, the eigenvalues of $I,X,Y,Z$ after QEC have the same form at the minimum, written as $27 (1 - p)^7 p^2 + 27 (1 - p)^6 p^3 + 99 (1 - p)^5 p^4 + 27 (1 - p)^4 p^5 + 57 (1 - p)^3 p^6 + 9 (1 - p)^2 p^7 + 9 (1 - p) p^8 + p^9$. It indicates that when p is above $0.950149$, the quantum channel can be purified.

The $5$-qubit code (II) in Ref.~\cite{Laflamme} and the $5$-qubit code (I) can transform to each other~\cite{DiVincenzo}, and the logical sates of the $5$-qubit code (II) can be written as
\be
|0_\mathcal{L}\rangle&=&\frac{1}{\sqrt{8}}[|00000\rangle-|10111\rangle-|01011\rangle+|11100\rangle\non\\
&&+|10010\rangle+|00101\rangle+|11001\rangle+|01110\rangle],\non
\ee
\be
|1_\mathcal{L}\rangle&=&\frac{1}{\sqrt{8}}[|11111\rangle-|01000\rangle+|10100\rangle-|00011\rangle\non\\
&&+|01101\rangle+|11010\rangle-|00110\rangle-|10001\rangle].\non
\ee
For general Pauli channel $\{\lambda_{II}=p_1,\lambda_{XX}=p_2,\lambda_{YY}=p_3,\lambda_{ZZ}=p_4,p_1+p_2+p_3+p_4=1\}$, the set of correctable errors consists of the identity operator $I$, and all the weight-one Pauli operators. After one-level QEC, the effective channel is obtained
\be
\label{e5}
\tilde{\lambda}=\{
\tilde{\lambda}_{II}=p^5_1+p^5_4+5p^4_1(p_2+p_3+p_4)+\non\\
p^4_4(p_1+p_2+p_3)+O(p^{i,i=0,1,2}_1)+O(p^{i,i=0,1,2,3}_4),\non\\
\tilde{\lambda}_{XX}=4p^4_2p_3+O(p^{i,i=0,1,2,3}_2),\non\\
\tilde{\lambda}_{YY}=p^5_2+p^5_3+5p^4_3(p_1+p_2+p_4)+\non\\
p^4_2(p_1+p_3+p_4)+O(p^{i,i=0,1,2}_3)+O(p^{i,i=0,1,2,3}_2),\non\\
\tilde{\lambda}_{ZZ}=4p^4_4p_1+O(p^{i,i=0,1,2,3}_4).
\ee
Here, $O(p^{i}_j)$ is the low-order term of $p^{i}_j$ ($j=1,2,3,4,i=0,1/0,1,2,3$). For the 5-qubit code, there is only one choice of correctable errors, which result in similar coefficient items for $I\&Y$ or $X\&Z$. Meanwhile, we notice that the eigenvalue of $I$ is mainly determined by the initial eigenvalues of $I\&Z$, and the eigenvalue of $Y$ is mainly determined by the initial eigenvalues of $Y\&X$. The eigenvalues of $X\&Z$ are minor terms, which may result from the structure of the $5$-qubit code. Taking only the dominant terms when $p_j$ is close to $1$, the eigenvalues of $I\&Y$ have the same form at the minimum, written as $5 (1 - p) p^4 + p^5$ (the initial channels are $I$-type and $Y$-type), or written as $2 (1 - p)^3 p^2 + 2 (1 - p)^2 p^3 + (1 - p) p^4 + p^5$ (the initial channels are $X$-type and $Z$-type). It indicates that when p is above $0.868877$, the $I$-type and $Y$-type quantum channel can be purified. When p is above $0.964831$, the $Z$-type quantum channel can be purified and changed to an $I$-type quantum channel, and the $X$-type quantum channel can be purified and changed to a $Y$-type quantum channel.

On the other hand, we also considered the amplitude damping channel and arbitrary numerical quantum channels. Because these noise channel cannot be accurately calculated as Pauli channels, we carried out some numerical simulations, in which the amplitude damping channel and arbitrary numerical quantum channels are randomly generated for observing evolution under QEC, and the result is similar to the Pauli channels'. Specifically, based on the performance of QEC for the amplitude damping channel, it can be classified as $I$-type channel when channel fidelity is above error-correction threshold, or $T$-type channel when channel fidelity is below error-correction threshold.

\section{Analysis of results}
\label{iii}

The results indicate that QEC can be used to classify and purify the independent quantum channel, and this conclusion is valid for the common error-correction codes. There is different purification threshold for different cases, and the specific effect of purification is depending on the structure of the code. For the $7$-qubit Steane code and the $5$-qubit code (I), the purification threshold is $0.935404$ and $0.868877$, respectively. For the $9$-qubit codes, the purification threshold is $0.950149$, and we should notice that the eigenvalue of $X$ and $Z$ will exchange after QEC with Shor code. For the codes above, the purification threshold is fixed, because the channel type can be exchanged through logical $X,Y,Z$ operators. But for the $5$-qubit code (II), the purification threshold is $0.868877$ (the initial channel are $I$-type and $Y$-type) or $0.964831$ (the initial channel are $X$-type and $Z$-type), and the $X$-type channel will be purified and changed to the $Y$-type channel, the $Z$-type channel will be purified and changed to the $I$-type channel. Because when encoded with this $5$-qubit code, $I$-type \& $Y$-type or $X$-type \& $Z$-type can be exchanged through logical operators, $Z$-type just can be partially converted to $I$-type through logical operators, and $X$-type just can be partially converted to $Y$-type through logical operators. Here, the purification threshold of code is obtained when considering the worst case, for one specific channel, the purification threshold may be below the purification threshold of the code.

In Eq.~(\ref{e1}) -~(\ref{e5}), the identification of the independent Pauli channels by QEC is demonstrated. Based on the results, we consider that the decoherence of quantum state may be related to the degree of identification for the state under quantum noise, and the results confirmed that the degree of purity of quantum channel determines its ability to retain the quantum property of the quantum state, not the fidelity. Because when one of $p_1, p_2, p_3, p_4$ is above some thresholds, the channel will have enough high degree of identification in QEC, to obtain one effective channel with the higher degree of identification. Meanwhile, the quantum state under the noise channel will have enough degree of identification to retain the quantum information. We should notice that the QEC scheme has actually been followed the rule without additional operations.

The results have a certain enlightening effect on realizing the logical Pauli operators in the QEC scheme. There are two methods to realize the logical Pauli operators in QEC, one is applying Pauli operators on the main system qubit before encoding, which will introduce correlated noise,
\be
\label{e6}
\tilde{P}_{S}=Tr_{A}[U^{\dag}[\Lambda[U[\rho_{A}\otimes P(\rho_{S})]]]].
\ee
Here, $\tilde{P}_{S} $ is the logical Pauli operator. $A$ is the assisted system, and $S$ is the main system qubit. $P$ is the independent Pauli operator act on main system qubit, and $\Lambda$ represents independent noise evolution.
The other one is applying logical Pauli operators after encoding or before decoding, which will introduce independent noise,
\be
\label{e7}
\tilde{P}_{S}=Tr_{A}[U^{\dag}[\Lambda[\tilde{P}[U(\rho_{A}\otimes\rho_{S})]]]].
\ee
In addition, $\tilde{P}$ is the independent Pauli operator act on all qubits, $U$ and $U^{\dag}$ are the encoding and decoding unitary operators. If the independent Pauli operators were ideal, and the channel fidelity of the independent noise is $p$, the fidelity of the logical Pauli operator $F(p)$ can be obtained through Eq.~(\ref{e1}) -~(\ref{e5}). When the independent Pauli operators are not ideal, and set the channel fidelity is $q$, the fidelity of the logical Pauli operator from Eq.~(\ref{e6}) is
\be
\label{e8}
F(p,q)_{1}>\approx q\times F(p).
\ee
And the fidelity of the logical Pauli operator from Eq. (\ref{e7}) is
\be
\label{e9}
F(p,q)_{2}<q\times F(p), p\&q  \, are \, discrete, \non\\
F(p,q)_{2}\geq q\times F(p), p\&q  \, are \, adjacent.
\ee
Here, `discrete' means the value gap between $p$ and $q$ is considerable, and `adjacent' means the value gap between $p$ and $q$ is negligible. Based on Eq. (\ref{e8}) and (\ref{e9}), we can choose a more appropriate method to implement the logical Pauli operator according to the actual situation. Meanwhile, we note that applying logical Pauli operators after encoding or before decoding have no influence on the fidelity of the logical Pauli operator in Eq. (\ref{e9}).

\section{Remarks and discussion}
\label{iv}

In previous works, QEC is one method for noise suppression, and is also used in fault-tolerant computation~\cite{A. Bolt,Theodore}. Based on this work, QEC can be used to classify and purify the independent quantum channel, which provides a method to trace the source of the quantum channel according to the effective channel. At the same time as purification for the quantum channel, $9$-qubit Shor code and $5$-qubit code (II) also realize some fixed logical gates. Specifically, the $9$-qubit Shor code will exchange the weight of $X$\&$Z$-items in the quantum channel. The $5$-qubit code (II) will suppress the weight of $X$\&$Z$ in the quantum channel, meanwhile, promotes the weight of $I$\&$Y$ in the quantum channel, which results in no $X$-type or $Z$-type quantum channel can exist after QEC with this code. From this work, we have a more clear cognition for QEC. For the unknown independent channel, QEC is not only one method to improve the fidelity, but also one method for identification, classification, and purification.


\begin{references}
\bibitem{Shor} P. W. Shor, Phys. Rev. A {\bf52}, R2493(R) (1995).
\bibitem{Steane} A. M. Steane, Phys. Rev. Lett. {\bf77}, 793 (1996).
\bibitem{Bennett} C. H. Bennett, D. P. DiVincenzo, J. A. Smolin, and W. K. Wootters, Phys. Rev. A {\bf54}, 3824 (1996).
\bibitem{Laflamme} R. Laflamme, C. Miquel, J. P. Paz, and W. H. Zurek, Phys. Rev. Lett. {\bf77}, 198 (1996).
\bibitem{KandL} E. Knill and R. Laflamme, Phys. Rev. A {\bf55}, 900 (1997).
\bibitem{Duan} L.-M. Duan and G.-C. Guo, Phys. Rev. Lett. {\bf79}, 1953 (1997).
\bibitem{Lidar} D. A. Lidar, I. L. Chuang, and K. B. Whaley, Phys. Rev. Lett. {\bf81}, 2594 (1998).
\bibitem{Zanardi} P. Zanardi and M. Rasetti, Phys. Rev. Lett. {\bf79}, 3306 (1997).
\bibitem{KandLV} E. Knill, R. Laflamme, and L. Viola, Phys. Rev. Lett. {\bf84}, 2525 (2000).
\bibitem{Zanardi2} P. Zanardi, Phys. Rev. A {\bf63}, 012301 (2000).
\bibitem{Kempe} J. Kempe, D. Bacon, D. A. Lidar, and K. B. Whaley, Phys. Rev. A {\bf63}, 042307 (2001).
\bibitem{Kribs} D. Kribs, R. Laflamme, and D. Poulin, Phys. Rev. Lett. {\bf94}, 180501 (2005).
\bibitem{Poulin 05} D. Poulin, Phys. Rev. Lett. {\bf95}, 230504 (2005).
\bibitem{Kribs2} D. W. Kribs and R. W. Spekkens, Phys. Rev. A {\bf74}, 042329 (2006).
\bibitem{Nielsen} M. A. Nielsen and I. L. Chuang, \emph{Quantum Computation and Quantum information}(Cambridge University Press,Cambridge, 2000).
\bibitem{Huang} L. Huang, B. You, X. H. Wu, and T. Zhou, Phys. Rev. A {\bf92}, 052320 (2015).
\bibitem{J. J. Wallman} J. J. Wallman and J. Emerson, Phys. Rev. A {\bf94}, 052325 (2016).
\bibitem{C.C2017} C. Chamberland, J. Wallman, S. Beale, and R. Laflamme, Phys. Rev. A {\bf95}, 042332 (2017).
\bibitem{S. J. Beale} S. J. Beale, J. J. Wallman, M. Guti\'{e}rrez, K. R. Brown, and R. Laflamme, Phys. Rev. Lett. {\bf121}, 190501 (2018).
\bibitem{E. Huang} E. Huang, A. C. Doherty, and S. Flammia, Phys. Rev. A {\bf99}, 022313 (2019).
\bibitem{Huang2} L. Huang, X. H. Wu, and T. Zhou, Phys. Rev. A  {\bf100}, 042321 (2019).
\bibitem{Gilchrist} A. Gilchrist, N. K. Langford, and M. A. Nielsen, Phys. Rev. A {\bf71}, 062310 (2005).
\bibitem{Emerson} J. Emerson, M. Silva, O. Moussa, C. Ryan, M. Laforest, J. Baugh, D. G. Cory, and R. Laflamme, Science {\bf317}, 1893 (2007).
\bibitem{Knill 08} E. Knill, D. Leibfried, R. Reichle, J. Britton, R. B. Blakestad, J. D. Jost, C. Langer, R. Ozeri, S. Seidelin, and D. J. Wineland, Phys. Rev. A {\bf77}, 012307 (2008).
\bibitem{Bendersky} A. Bendersky, F. Pastawski, and J. P. Paz, Phys. Rev. A {\bf80}, 032116 (2009).
\bibitem{Magesan 11} E. Magesan, J. M. Gambetta, and J. Emerson, Phys. Rev. Lett. {\bf106}, 180504 (2011).
\bibitem{Magesan 12} E. Magesan, J. M. Gambetta, B. R. Johnson, C. A. Ryan, J. M. Chow, S. T. Merkel, M. P. da Silva, G. A. Keefe, M. B. Rothwell, and T. A. Ohki et al., Phys. Rev. Lett. {\bf109}, 080505 (2012).
\bibitem{X.-H. Wu} X.-H. Wu and K. Xu, Quantum Inf. Proc. {\bf12}, 1379 (2013).
\bibitem{Huang3} L. Huang, X. H. Wu, and T. Zhou, arXive e-print quant-ph/2110.05130 (2021).
\bibitem{DiVincenzo} D. P. DiVincenzo and P. W. Shor, Phys. Rev. Lett. {\bf77}, 3260 (1996).
\bibitem{A. Bolt} A. Bolt, G. Duclos-Cianci, D. Poulin, and T. M. Stace, Phys. Rev. Lett. {\bf117}, 070501 (2016).
\bibitem{Theodore} T. J. Yoder, R. Takagi, and I. L. Chuang, Phys. Rev. X {\bf6}, 031039 (2016).
\end{references}
\end{document}